\documentclass[a4paper,fleqn,usenatbib]{mnras}

\usepackage{graphicx}

\usepackage{hyperref}
\usepackage{amsmath}
\usepackage{amssymb}
\usepackage{wasysym}
\usepackage{color}
\usepackage{pifont}
\usepackage{multirow}
\usepackage{braket}
\AtBeginDocument{\renewcommand{\d}{\textrm{d}}}

\newcommand{\cmark}{\ding{51}}%
\newcommand{\xmark}{\ding{55}}%

\title[Anisotropy of \textit{Fermi}/GBM gamma-ray bursts]{Testing the anisotropy in the angular distribution of \textit{Fermi}/GBM gamma-ray bursts}
\author[M. Tarnopolski]{M. Tarnopolski\thanks{E-mail:
mariusz.tarnopolski@uj.edu.pl}\\
Astronomical Observatory, Jagiellonian University, Orla 171, 30-244 Krak\'{o}w, Poland}

\date{Accepted XXX. Received YYY; in original form ZZZ}
\pubyear{2017}

\begin{document}
\label{firstpage}
\pagerange{\pageref{firstpage}--\pageref{lastpage}}
\maketitle

\begin{abstract}
Gamma-ray bursts (GRBs) were confirmed to be of extragalactic origin due to their isotropic angular distribution, combined with the fact that they exhibited an intensity distribution that deviated strongly from the $-3/2$ power law. This finding was later confirmed with the first redshift, equal to at least $z=0.835$, measured for GRB970508. Despite this result, the data from {\it CGRO}/BATSE and {\it Swift}/BAT indicate that long GRBs are indeed distributed isotropically, but the distribution of short GRBs is anisotropic. {\it Fermi}/GBM has detected 1669 GRBs up to date, and their sky distribution is examined in this paper. A number of statistical tests is applied: nearest neighbour analysis, fractal dimension, dipole and quadrupole moments of the distribution function decomposed into spherical harmonics, binomial test, and the two point angular correlation function. Monte Carlo benchmark testing of each test is performed in order to evaluate its reliability. It is found that short GRBs are distributed anisotropically on the sky, and long ones have an isotropic distribution. The probability that these results are not a chance occurence is equal to at least 99.98\% and 30.68\% for short and long GRBs, respectively. The cosmological context of this finding and its relation to large-scale structures is discussed.
\end{abstract}


\section{Introduction}
\label{intro}
Gamma-ray bursts (GRBs) are highly energetic events that often are brighter than all other $\gamma$-ray objects visible on the sky combined, with an emission peak in the 200--500 keV region \citep[for recent reviews, see][]{nakar,zhang4,gehrels2,berger,rees,dainotti}. \citet{mazets} first observed a bimodal distribution of $T_b$ (taken to be the time interval within which fall $80\!-\!90\%$ of the measured GRB's intensity) drawn for 143 events observed in the KONUS experiment. {\it CGRO}/BATSE 1B data release \citep{meegan92} was followed by further investigation of the $T_{90}$ (time between the 5\% and 95\% of the energy release in the GRB prompt phase) distribution of 222 GRBs \citep{kouve} that lead to establishing the common classification of GRBs into short ($T_{90}<2\,{\rm s}$) and long ($T_{90}>2\,{\rm s}$), and based on which GRBs are most commonly classified \citep[but see also][]{kann,bromberg,tarnopolski15a,li2}. The progenitors of long GRBs are associated with supernovae related with collapse of massive stars \citep{woosley}. Progenitors of short GRBs are thought to be NS-NS or NS-BH mergers \citep{nakar}, and no connection between short GRBs and supernovae has been proven \citep{zhang5}. Indeed, \citet{balazs03} showed that the relation between fluence and duration is different for short and long {\it CGRO}/BATSE GRBs, indicating an intrinsic difference in the physical mechanisms of energy release between the two classes.

The existence of an intermediate-duration GRB class, consisting of GRBs with $T_{90}$ in the range $2\!-\!10\,{\rm s}$, was put forward based on the analysis of {\it CGRO}/BATSE 3B data, consisting of 797 GRBs \citep{horvath98,mukh}. It was supported with the use of the more complete {\it CGRO}/BATSE dataset, containing 1929 GRBs \citep{horvath02}. Evidence for a third normal component in the $\log T_{90}$ distribution was also found in {\it Swift}/BAT data (\citealt{horvath08,zhang2,huja,horvath10}; see also \citealt{veres}) and in the {\it RHESSI} database \citep{ripa1,ripa2,ripa3}. {\it BeppoSAX} dataset was shown to be in agreement with earlier results regarding the bimodal distribution, and the detection of an intermediate-duration component in the GRB population was established on a lower, compared to {\it CGRO}/BATSE and {\it Swift}/BAT, significance level \citep{horvath09}. It is important to note that in {\it BeppoSAX} only the components related to intermediate and long GRBs were detected, the short ones being absent. Interestingly, \citet{zitouni} re-examined the {\it CGRO}/BATSE Current Catalog as well as the {\it Swift}/BAT dataset, and found that a mixture of three Gaussian (3G) distributions fits the {\it Swift}/BAT data better than a two-component Gaussian mixture (2G), while in the {\it CGRO}/BATSE case statistical tests did not support the presence of a third component. Regarding {\it Fermi}/GBM, a 3G is a better fit than a 2G, however the presence of a third group in the $\log T_{90}$ distribution was found to be unlikely \citep{tarnopolski15b,tarnopolski16a}, and that a mixture of two intrinsically skewed distributions follows better the (bimodal) $\log T_{90}$ distribution than a 3G \citep{tarnopolski16a}.

The {\it Fermi}/GBM dataset \citep{gruber,kienlin} has been examined widely for other reasons, too. Its redshift distribution was investigated \citep{ackermann} confirming the observation that short GRBs have systematically lower redshifts than long ones. The Amati correlation was investigated \citep{basak,gruber2}, and a link between short and long GRBs was discovered \citep{muccino}. It gave insight into the GRB afterglow population \citep{racusin}, allowed to observe a number of high-energy GRBs (with photon energies exceeding 100~MeV, or even 10~GeV; \citealt{atwood}), and provided a verification of the short--long classification \citep{zhang3}.

After the early hypothesis of \citet{klebesadel} that GRBs are of extragalactic origin, \citet{colgate} interpreted GRBs as coming from supernovae at cosmological distances (10--30 Mpc, encompassing up to 5000 galaxies). In an Euclidean space, the intensity distribution is expected to be in the form of a $-3/2$ power law (e.g., \citealt{paczynski91}) in case of a uniform spatial distribution; however, \citet{usov} showed---within a Friedmann cosmology---that at cosmological distances the Euclidean approximation fails and the flux distribution is expected to deviate from the $-3/2$ power law \citep{meegan92,fishman}. Hence, for GRBs with high enough redshifts such a deviation need not indicate an inhomogeneous spatial distribution. Thus, \citet{paczynski86} predicted GRBs to reside at redshifts $z=1\!-\!2$. A detailed analytical treatment of the integral number count flux distribution was conducted within the Friedmann cosmology with fulfilled conditions for homogeneity and isotropy \citep{meszaros6,meszaros7,meszaros8}; this allowed to further confirm the extragalactic origin of GRBs, and a prediction that they can be observable up to redshift $z=20$ was put forward. To date, the furtherst GRB has a redshift of $z=9.4$ \citep{cucchiara}. The first redshift measurement, established for GRB970508, which with $0.835<z\lesssim 2.3$ was indeed placed at a cosmological distance\footnote{A comoving distance in a flat Friedmann cosmology \citep{carroll} with $H_0=67\,{\rm km\,s^{-1}\,Mpc^{-1}}$ and $\Omega_m=0.32$ \citep{planck}; these parameters are used throughout this work.} of at least 3 Gpc \citep{metzger}, corroborated the previous statistical inferences about the extragalactic nature of GRB in a direct manner.

Initially, the angular distribution of GRBs was found to be isotropic \citep{tegmark}. However, \citet{balazs} examined a sample of 2025 GRBs using the dipole-quadrupole and binomial tests, and found that short GRBs are distributed anisotropically, while long ones are isotropic. \citet{meszaros4} came to the same conclusion using the count-in-cells method. Surprisingly, \citet{meszaros3} found that in a bigger sample of GRBs both short and long ones are isotropic. Using the nearest neighbour analysis (NNA), it was verified, contrary to previous results, that the sample of short GRBs is isotropic, and long ones are anisotropically distributed \citep{meszaros03}. The anisotropy of short GRBs was again confirmed by \citet{maglio} by using the two point angular correlation function (2pACF) by means of the Hamilton estimator \citep{hamilton}. Contrary to this, \citet{bernui} claimed an intrinsic isotropy in a sample of short GRBs, performing a 2pACF and the sigma-map analysis. Finally, \citet{vavrek} confirmed that the short GRBs are distributed anisotropically by means of other statistical tests, i.e., Voronoi tesselation, minimal spanning tree, and multifractal spectra (see Table~\ref{tbl0} for a summary of the previous works). Cosmological consequences of an anisotropic distribution were discussed lately by \citet{meszaros,meszaros5}. With the growing number of GRBs with known redshift, another approach to the anisotropy problem was undertaken by \citet{horvath14}, who performed a two-dimensional Kolmogorov-Smirnov test and $k$-th nearest neighbour test on the 283 GRBs with known $z$, as detected by {\it Swift}/BAT, in different radial binnings. They found that a GRB sample of 31 objects in $1.6<z<2.1$ is significantly anisotropic. \citet{ukwatta} also studied the redshift- and duration-dependent clustering of {\it Swift}/BAT GRBs, using proximity measures and kernel density estimation, and utilising GRBs detected by {\it CGRO}/BATSE (2037 objects), {\it Swift}/BAT (889 objects) and {\it Fermi}/GBM (997 objects) they found marginal anisotropy for very short GRBs (with $T_{90}<100\,{\rm ms}$). Moreover, they argued that there is no evidence for anisotropy in an updated sample of 34 {\it Swift}/BAT GRBs in the same $z$ range that was examined by \citet{horvath14}.

\begin{table}
\begin{center}
\caption{Results of previous works on the celestial distribution of short and long GRBs. The checkmark (\cmark) denotes isotropy.}
\label{tbl0}
\begin{tabular}{ccc}
\hline
Reference & Short GRBs & Long GRBs \\
\hline
\citet{balazs}     & \xmark & \cmark \\
\citet{meszaros4}  & \xmark & \cmark \\
\citet{meszaros3}  & \cmark & \cmark \\
\citet{meszaros03} & \cmark & \xmark \\
\citet{maglio}     & \xmark & --- \\
\citet{bernui}     & \cmark & --- \\
\citet{vavrek}     & \xmark & --- \\
\hline
\end{tabular}
\end{center}
\end{table}

It is worth to note that according to some research the intermediate GRBs are distributed anisotropically on the sky \citep{meszaros4,meszaros3,vavrek}. On the other hand, \citet{meszaros03} argued that the intermediate class is isotropic. The origin and existence of the intermediate GRBs remain elusive as theoretical models still need to account for an apparent bimodality in duration distribution \citep{janiuk,nakar}. Finally, not only the existence of an intermediate class was investigated (and remains unsettled), but also subclass classifications of long GRBs were proposed \citep[e.g.,][]{gao}.

The {\it Fermi}/GBM data have not been examined yet with regard to their celestial isotropy, except for a smaller sample of 997 GRBs that were examined together with GRBs observed by other satellites \citep{ukwatta}, and as part of a sample of the 361 GRBs with known redshift \citep{balazs15}. The aim of this article is to study the celestial distribution of {\it Fermi}/GBM data. A number of statistical tests (NNA, fractal dimension, dipole-quadrupole test, binomial test, and the 2pACF) is applied herein to a sample of 1669 {\it Fermi}/GBM GRBs in order to verify whether the data are consistent with the previous findings among datasets from other satellites, i.e., whether the celestial distribution of short GRBs is anisotropic, and that of the long ones is isotropic.

This paper is organized in the following manner. Section~\ref{data} describes the GRB sample. In Sect.~\ref{methods} a detailed description of the methods is given. This consists of the presentation of the statistical tests, as well as the benchmark testing approach to each of them to ensure that the probability of a false detection is low enough. In Sect.~\ref{results} the results of these tests are presented. Section~\ref{disc} is devoted to discussion, and is followed by concluding remarks gathered in Sect.~\ref{conc}.

\section{Dataset}
\label{data}
\begin{figure*}
\centering
\includegraphics[width=0.8\textwidth]{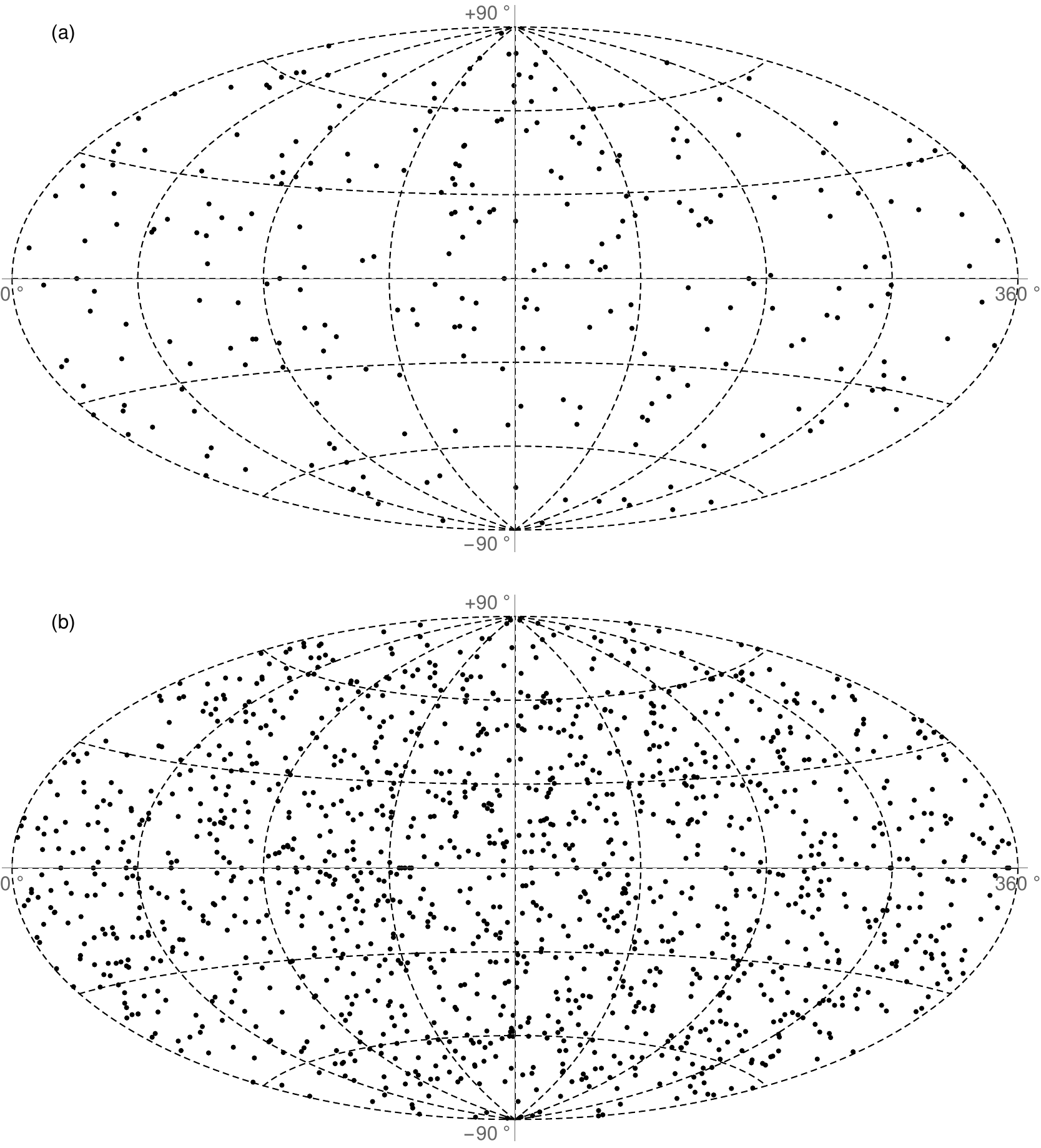}
\caption{The sky distribution of (a) short and (b) long GRBs in Galactic coordinates in Aitoff projection.}
\label{fig1}
\end{figure*}

The dataset from {\it Fermi}/GBM\footnote{\url{http://heasarc.gsfc.nasa.gov/W3Browse/fermi/fermigbrst.html}, accessed on July 29, 2015.} \citep{gruber,kienlin} is examined in this paper. The considered data set consists of the celestial locations, given in Galactic coordinates $(b,l)$, and the durations $T_{90}$ for each of the 1669 GRBs. Four of them do not have a reported $T_{90}$, and among the remaining ones 278 are short ($T_{90}<2\,{\rm s}$), and 1387 are long ($T_{90}>2\,{\rm s}$). Note that the ratio of short-to-long GRBs is 1:5, being in between the {\it CGRO}/BATSE and {\it Swift}/BAT ratios, which are 1:3 and 1:9, respectively. Despite that it was shown that for the GRBs observed by {\it Fermi}/GBM a limit of $T_{90}=2.05\,{\rm s}$ is more appropriate \citep{tarnopolski15a}, the difference between the new limit and the commonly applied limit of $2\,{\rm s}$ is of negligible significance, therefore the conventional $2\,{\rm s}$ limit \citep{kouve} is applied herein. The sky distributions of short and long GRBs are shown in Fig.~\ref{fig1}. The data are divided into three samples: short GRBs, long ones, and a joined sample of all GRBs. The latter consists of short, long, and the four undefined GRBs. The elusive intermediate class is not examined herein, as it was shown that their presence in the {\it Fermi}/GBM Catalog is unlikely \citep{tarnopolski15b,tarnopolski16a}. A computer algebra system {\sc Mathematica} is used throughout this paper, with the following exceptions: the fractal dimension is computed with a simple {\sc Python} code, and for computation of the 2pACF a {\sc C++} programme is used.

\section{Methods}
\label{methods}
A repeating key concept in the majority of the statistical tests performed is the measurement of the angular distance between two GRBs observed on the sky. Given the Galactic coordinates of the bursts $(b_1,l_1)$ and $(b_2,l_2)$, where $b\in[-90^\circ,90^\circ]$ is the latitude, and $l\in[0^\circ,360^\circ)$ is the longitude, this distance is calculated as
\begin{equation}
\Delta\sigma_{12}=\arccos\left(\sin b_1\sin b_2+\cos b_1\cos b_2\cos\Delta l\right),
\label{eq1}
\end{equation}
where $\Delta l=l_1-l_2$. Throughout this paper, whenever a phrase {\it distance between GRBs} is used, it should be understood in the sense of the angular distance given by Eq.~(\ref{eq1}).

For benchmark testing, it will be necessary to randomly generate a set of points coming from a population isotropically distributed on the sky\footnote{This is also a part of the algorithm to compute the 2pACF for the observed GRBs; see Sect.~\ref{m9}.}. This is achieved with the following formulae:
\begin{subequations}
\begin{align}
b &= \arcsin(2u-1), \label{eq2a} \\
l &= 2\pi v, \label{eq2b}
\end{align}
\label{eq2}%
\end{subequations}
where $u,v$ are independent random variables distributed uniformly in $[0,1)$.

The benchmark testing is especially desired for short GRBs. The detected anisotropy in this case may be questioned due to the smallness of the sample; hence, it is important to estimate the significance of the detection. In all tests, a significance level of $\alpha=0.05$ is applied. The tests are recapitulated in the following sections.

\subsection{Nearest neighbour analysis (NNA)}
\label{m1}
The theory of NNA was formulated by \citet{scott} (see also \citealt{slechta}), and specified for the sky distribution of GRBs by \citet{meszaros03}.

Assume there are $N$ points (GRBs) distributed randomly on a sphere (the sky). Let us choose one point and consider a ring with infinitesimal thickness $\d\beta$, located at an angular distance $\beta$ from the chosen point; hence, this ring is defined by the interval $[\beta,\beta+\d\beta]$. The probability of having $L-1$ points at a distance smaller than $\beta$, and exactly one point inside this ring, is given by
\begin{equation}
\begin{array}{l}
p_L(\beta) = \frac{(N-1)!}{2^{N-1}(N-L-1)!(L-1)!} \\
\textcolor{white}{p_L(\beta) = }\times \sin\beta(1-\cos\beta)^{L-1}(1+\cos\beta)^{N-L-1},
\end{array}
\label{eq3}
\end{equation}
where $L=1,2,\ldots,N-1$. For $L=1$ the (first) nearest neighbour lies in the ring, for $L=2$ only the second nearest neighbour lies in the ring, and so on. The probability that there are exactly $L$ points in the neighbourhood (defined as the distances smaller than $\beta$) of the chosen point is given by the integral 
\begin{equation}
P_L(\beta)=\int\limits_0^\beta p_L(\beta')\d\beta',
\label{eq4}
\end{equation}
which for $L=1$ and $L=2$ yields
\begin{equation}
P_1(X_1)=1-\left(1-\frac{X_1}{N-1}\right)^{N-1}
\label{eq5}
\end{equation}
and
\begin{equation}
\begin{array}{l}
P_2(X_2) = 1-\left(1-\frac{X_2}{(N-1)(N-2)}\right)^{N-1} \\
\textcolor{white}{P_2(X_2) = } - \frac{X_2}{N-1}\left(1-\frac{X_2}{(N-1)(N-2)}\right)^{N-2},
\end{array}
\label{eq6}
\end{equation}
respectively, with \mbox{$X_1=(N-1)\sin^2\frac{\beta}{2}$} and \mbox{$X_2=(N-1)(N-2)\sin^2\frac{\beta}{2}$}. The function $P_1(X_1)$ is defined for $0\leq X_1\leq N-1$, and $P_2(X_2)$ is defined for $0\leq X_2\leq (N-1)(N-2)$. In practice, they nearly plateau for $X_1\ll N-1$ and $X_2\ll (N-1)(N-2)$.

The integral probabilities for different $L$ are the theoretical cumulative distribution functions (CDFs). Computing the empirical CDFs from the observed angular distribution of GRBs, one may perform the Kolmogorov-Smirnov test\footnote{Using {\sc Mathematica}'s built-in \texttt{KolmogorovSmirnovTest} command in order to retrieve the $p$-value.} \citep{kolmogorov} to verify whether the celestial locations of GRBs come from an isotropic population or not. To perform the test for $L=1$, the empirical CDF of the first nearest neighbour distance should be computed according to Eq.~(\ref{eq1}); for $L=2$, the second nearest neighbour has to be found. Then, the maximal absolute difference $D$ between the empirical and theoretical CDFs needs to be calculated, and compared with the critical value $D_{\rm crit}$ for given $\alpha$ and $N$. The null hypothesis that the examined dataset comes from an isotropic distribution cannot be rejected if $D<D_{\rm crit}$, or equivalently, if the $p$-value exceeds the significance level $\alpha$. For $N>40$, usually an approximation is used \citep[][pp. 427-431]{sachs}:
\begin{equation}
D_{\rm crit}=\sqrt{\frac{-0.5\ln\left(\frac{\alpha}{2}\right)}{N}}.
\label{eq6a}
\end{equation}

In general, one can perform tests for $N-1$ different values of $L$. However, this is not necessary. If the null hypothesis that the GRBs are located isotropically on the sky is true, none of the tests should reject this hypothesis. Increasing $L$ is unlikely to produce any result different than for smaller values. This is because among the $N(N-1)/2$ distances between all GRBs, $2N-3$ are independent \citep[for a detailed discussion, see][]{meszaros03}. Using $L=1$ and $L=2$, one exploits $2N$ distances, which is nearly identical to the number of independent distances. Therefore, only those two $L$ values are tested herein.

\subsection{NNA benchmark testing}
\label{m2}
In order to evaluate the reliability of the NNA, 1000 Monte Carlo (MC) realisations of the isotropic sky distribution are generated according to Eq.~(\ref{eq2}) for short (278 points), long (1387 points) and a joined sample of 1669 points. For each realisation, the empirical CDF of the first ($L=1$) and second ($L=2)$ nearest neighbour distance is computed, and compared with the theoretical CDFs given by Eq.~{(\ref{eq5}) and (\ref{eq6}). Next, the Kolmogorov-Smirnov test is performed. The maximal absolute difference $D$ between the two CDFs is then compared with the critical value $D_{\rm crit}$ from Eq.~(\ref{eq6a}). If $D>D_{\rm crit}$, then the null hypothesis that the data are isotropically distributed is rejected. This is a false rejection, as the simulated data come from an isotropic distribution. These instances will be referred to as {\it false-anisotropic}.

The results of the benchmark testing are gathered in Table~\ref{tbl1}. The number of MC realisations with $N=278$ (corresponding to the sample of short GRBs) were correctly classified as isotropic in 717 cases. It follows that there is a probability of 0.283 of a false-anisotropic detection. For $N=1387$, corresponding to long GRBs, the probability of correctly inferring the isotropy of a dataset is 0.832. Among the isotropic samples with $N=1669$ (all GRBs) the ratio of correct classifications is 0.822. Note that for $N=1387$ and $N=1669$, the NNA with $L=2$ did not make even a single false detection, hence it is more sensitive to the underlying isotropy than $L=1$.
\begin{table}
\begin{center}
\caption{Results of the NNA benchmark testing for 1000 MC realisations of short, long and all GRBs. The first row denotes the number of realisations that were correctly classified as isotropic; the second row gives the number of sets which isotropy was rejected based on the Kolmogorov-Smirnov test for $P_1(X_1)$; the third row -- similarly for $P_2(X_2)$; the fourth row gives the number of realisations whose isotropy was rejected based on the Kolmogorov-Smirnov test for both $P_1(X_1)$ and $P_2(X_2)$. The rows from second to fourth combined are the false-anisotropic instances.}
\label{tbl1}
\begin{tabular}{cccc}
\hline
 & short & long & all \\
\hline
isotropic   & 717 & 832 & 822 \\
$L=1$       & 107 & 168 & 178 \\
$L=2$       & 124 & 0   & 0   \\
$L=1,\,L=2$ & 52  & 0   & 0   \\
\hline
\end{tabular}
\end{center}
\end{table}

\subsection{Fractal dimension}
\label{m3}
The fractal dimension is a measure of how much space is covered by a set. It is often measured with the correlation dimension, $d_C$ \citep{grassberger,grassberger2,alligood,ott}, which takes into account the local densities of the points in the examined dataset. For usual 1-D, 2-D or 3-D cases (those include a coverage by isotropically distributed points) the $d_C$ is equal to 1, 2 and 3, respectively. Usually, a fractional correlation dimension is obtained for fractals \citep{mandelbrot}. However, when one deals with clustering, the correlation dimension will differ from the integer value corresponding to the dimension of the embedding space. Hence, the correlation dimension might serve as an indicator of anisotropy in the celestial distribution of GRBs \citep[see also][]{yadav}.

The correlation dimension is defined as
\begin{equation}
d_C=\lim_{r\rightarrow 0}\frac{\ln C(r)}{\ln r},
\label{eq7}
\end{equation}
with the estimate for the correlation function $C(r)$ as
\begin{equation}
C(r)=\frac{2}{N(N-1)}\sum_{i=1}^N\sum_{j=i+1}^N \Theta(r-\Delta\sigma_{i,j}),
\label{eq8}
\end{equation}
where the Heaviside step function $\Theta$ adds to $C(r)$ only points $(b_i,l_i)$ in an angular distance smaller than $r$ from $(b_j,l_j)$ and vice versa. The limit in Eq.~(\ref{eq7}) is attained by using multiple values of $r$ and fitting a straight line to the linear part of the dependencies obtained. The correlation dimension is estimated as the slope of the linear regression. In case of an isotropic celestial distribution, the expected $d_C$ is equal to 2, and a significant deviation from $d_C=2$ means that the GRBs are distributed anisotropically. The fractal dimension cannot be greater than 2 for a distribution on a surface (within the measurement errors), so a significant deviation toward $d_C<2$ is expected for anisotropic datasets. Nevertheless, one might expect the occurence of values slightly greater than 2 due to statistical fluctuations.

\subsection{Fractal dimension benchmark testing}
\label{m4}
For an isotropically distributed sample, the expected correlation dimension is $d_C=2$. However, due to the sample's finiteness, the computed dimension might deviate from the expected value \citep{tarnopolski14}. To quantify what deviation can be ascribed to statistical fluctuations, 1000 MC realisations are generated according to Eq.~(\ref{eq2}) for short (278 points), long (1387 points), and a joined sample of 1669 points. For each realisation, the correlation dimension is calculated\footnote{Using a simple parallel {\sc Python} programme, based on the one in \citep{tarnopolski14}; available on request.} according to Eq.~(\ref{eq7}) and (\ref{eq8}). Due to different $N$, the region of a linear fit was different for each sample size, and was chosen based on the results for real GRBs (see Sect.~\ref{r2}). The mean and standard deviation for each $N$ are computed and gathered in Table~\ref{tbl2}.
\begin{table}
\begin{center}
\caption{Mean correlation dimensions of 1000 MC realisations of short, long and all GRBs in the benchmark testing.}
\label{tbl2}
\begin{tabular}{cccccc}
\hline
Sample & $N$ & Mean $d_C$ & \begin{tabular}[c]{@{}c@{}}Standard\\deviation\end{tabular} & \begin{tabular}[c]{@{}c@{}}Fitting\\region, $\ln r$\end{tabular} \\
\hline
short & 278  & 1.966 & 0.035 & $(-2,0)$ \\
long  & 1387 & 1.982 & 0.012 & $(-3,0)$ \\
all   & 1669 & 1.990 & 0.016 & $(-4,0)$ \\
\hline
\end{tabular}
\end{center}
\end{table}
The mean correlation dimensions are all very close to 2. However, as displayed in Fig.~\ref{fig2}, due to a relatively wide range of the dimensions obtained, one cannot simply establish the validity of the detection of anisotropy judging only on the displacement from $d_C=2$. Instead, a probability {\it via} a $p$-value will be reported. A $p$-value will be calculated as the probability of finding, among isotropic datasets of length $N$, a correlation dimension deviating from $d_C=2$ at least as much as the observed one. This will be a two-sided $p$-value, and a probability of a deviation from $d_C=2$ toward smaller values at least as much as the observed one will be a one-sided $p$-value.
\begin{figure}
\centering
\includegraphics[width=\columnwidth]{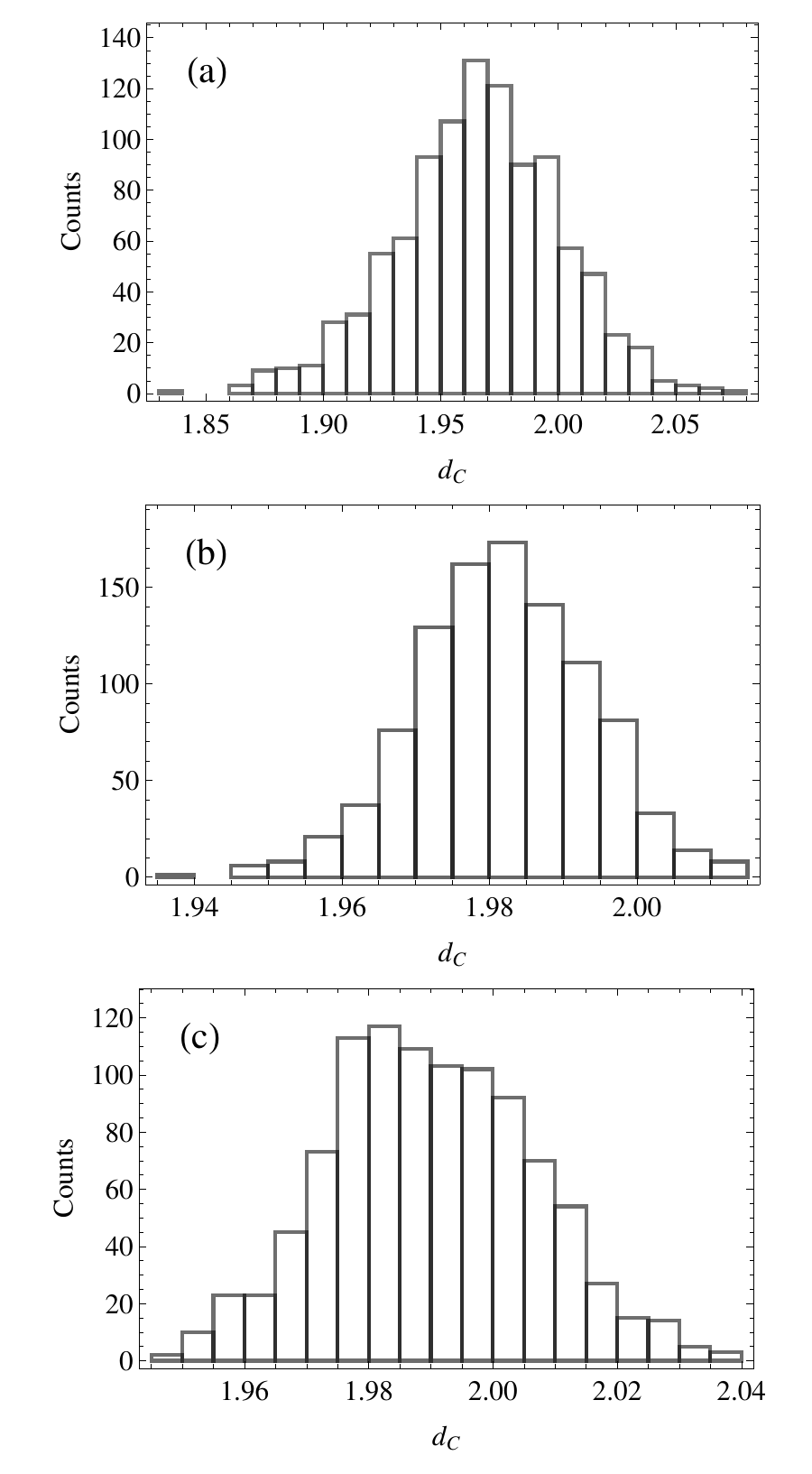}
\caption{The distributions of the correlation dimension of 1000 MC realisations for (a) short, (b) long, and (c) all GRBs. Note the dispersion in (a) is about 2--3 times larger than in (b) and (c), what is caused by the smallness of the sample.}
\label{fig2}
\end{figure}

\subsection{Dipole-quadrupole test}
\label{m5}
If the GRBs are located isotropically on the sky, then their angular distribution function $\omega$ is independent on their physical properties, and may depend only on the celestial coordinates, hence $\omega=\omega(b,l)$. Moreover, isotropy means that $\omega(b,l)=1/(4\pi)$, therefore the probability of finding a GRB in a solid angle $\Omega$ is equal to $\Omega/(4\pi)$, and is independent on its location on the sky. These observations naturally lead to a decomposition of $\omega(b,l)$ into spherical harmonics in the following manner \citep{balazs}:
\begin{equation}
\begin{array}{l}
\omega(b,l) = \sum\limits_{i,j\in\mathbb{Z}}^N c_{i,j}\omega_{i,j}Y_{i,j}(b,l) = \frac{1}{\sqrt{4\pi}}\omega_{0,0} \\
-\sqrt{\frac{3}{4\pi}} \big( \omega_{1,-1}\cos b\sin l-\omega_{1,1}\cos b\cos l+\omega_{1,0}\sin b \big) \\
+\sqrt{\frac{15}{16\pi}} \big( \omega_{2,-2}\cos^2 b\sin 2l+\omega_{2,2}\cos^2 b\cos 2l \\
\textcolor{white}{+\sqrt{\frac{15}{16\pi}}} -\omega_{2,-1}\sin 2b\sin l-\omega_{2,1}\sin 2b\cos l \big) \\
+\sqrt{\frac{5}{16\pi}}\omega_{2,0} \big( 3\sin^2 b-1 \big)+\ldots
\end{array}
\label{eq9}
\end{equation}
where $c_{i,j}$ are the numerical coefficients, $Y_{i,j}(b,l)$ are the trigonometric terms dependent on $b$ and $l$, and higher order harmonics are omitted. In Eq.~(\ref{eq9}), the first term is a monopole term, the next three ones are the dipole terms, and the following five ones are the quadrupole terms. Because the spherical harmonics are orthogonal functions, the $\omega_{i,j}$ coefficients are given by the functional scalar products. Isotropy means that $\omega_{i,j}=0$ for $i\neq 0,\,j\neq 0$ (leaving only $\omega_{0.0}=1$). Due to $\omega(b,l)$ being given in discrete points, the integral scalar product is transformed into a sum over all observed GRBs, and the isotropy criterion is
\begin{equation}
\omega_{i,j}\propto\frac{1}{N}\sum\limits_{k=1}^N Y_{i,j}(b_k,l_k)=0,
\label{eq10}
\end{equation}
i.e., the expected value (mean) of the coefficients $\omega_{i,j}$ (with $i\neq 0,j\neq 0$) is equal to zero \citep[for further details, see][]{balazs}. In order to test their zero-mean, a Student $t$ test \citep{student} will be performed\footnote{Using {\sc Mathematica}'s built-in \texttt{TTest} command.}. If any of the $\omega_{i,j}$ coefficients turns out to be statistically different from zero, an anisotropy in the sky distribution of GRBs will be claimed.

\subsection{Dipole-quadrupole benchmark testing}
\label{m6}
In order to test the reliability of the dipole-quadrupole test, 1000 MC realisations are generated according to Eq.~(\ref{eq2}) for short (278 points), long (1387 points), and a joined sample of 1669 points. For each realisation, the $\omega_{i,j}$ coefficients are computed according to Eq.~(\ref{eq10}), and tested whether they are equal to zero or not with the Student $t$ test. The results are gathered in Table~\ref{tbl3}. For all three sample sizes, about $1/3$ of the sets were reported to be anisotropic, despite coming from an isotropic population. This means that the possibility of a false detection of anisotropy is relatively high. In these false-anisotropic cases, on average one $\omega_{i,j}$ coefficient was statistically non-zero (although sometimes it was more than one). The distribution of non-zero coefficients is more or less uniform, and they appear approximatelly 50 times for each sample size each. Hence, the dipole-quadrupole test is not much trustworthy by itself, but may be still used to verify if it is consistent with the outcomes of other tests.
\begin{table}
\begin{center}
\caption{The number of false anisotropic detections in 1000 MC realisations of short, long and all GRBs in the dipole-quadrupole benchmark test. Note that the total number of false-anisotropic detections in not the sum of the corresponding columns (see text).}
\label{tbl3}
\begin{tabular}{ccccc}
\hline
$\omega_{i,j}$ & short & long & all \\
\hline
$\omega_{1,-1}$ & 46 & 45 & 41 \\
$\omega_{1,0}$  & 47 & 50 & 44 \\
$\omega_{1,1}$  & 53 & 55 & 51 \\
$\omega_{2,-2}$ & 59 & 46 & 47 \\
$\omega_{2,-1}$ & 49 & 50 & 50 \\
$\omega_{2,0}$  & 59 & 49 & 46 \\
$\omega_{2,1}$  & 53 & 59 & 50 \\
$\omega_{2,2}$  & 38 & 39 & 54 \\
\hline
\begin{tabular}[c]{@{}c@{}}False\\anisotropic\end{tabular} & 337 & 334 & 324 \\
\hline
\end{tabular}
\end{center}
\end{table}

\subsection{Binomial test}
\label{m7}
The probability $p$ of finding a GRB in a solid angle $\Omega$ is $p=\Omega/(4\pi)$, and the probability of finding it outside is $1-p$. Hence, the probability of having $k$ GRBs in $\Omega$ is given by the binomial (Bernoulli) distribution (\citealt{kendall,meszaros97} and references therein):
\begin{equation}
P_p(N,k)={N \choose k}p^k(1-p)^{N-k},
\label{eq11}
\end{equation}
with the expected (mean) value equal to $Np$. Assume that $k_{\rm obs}$ is the observed number of GRBs in a solid angle $\Omega$. It will be tested whether it is consistent with the null hypothesis that the distribution is isotropic. Certainly, any $0\leq k_{\rm obs}\leq N$ may occur, but if its probability is small enough it is unlikely that the underlying distribution is isotropic.

Consider the deviation of $k_{\rm obs}$ from the mean $Np$, given by $k_0=|k_{\rm obs}-Np|$. Then, one may introduce a probability of a deviation not smaller than $k_{\rm obs}$ from the mean $Np$ as
\begin{equation}
\begin{array}{l}
P(N,k_{\rm obs},p)=P(Np+k_0\leq X\leq Np-k_0) \\
\quad=1-P(X\leq Np+k_0)+P(X\leq Np-k_0),
\end{array}
\label{eq12}
\end{equation}
where $P(X\leq X_0)$ is the CDF of a binomial distribution \citep{balazs}. If this probability is high, then the dataset may come from an isotropic population; if it is small, the isotropy is unlikely and the set is probably anisotropic.

To perform a binomial test, the sky will be divided into two disjoint parts (neither of these parts must be a connected compact region), and in general the division may be arbitrary.

\subsection{Binomial benchmark testing}
\label{m8}
To divide the sky into two regions for the benchmark testing, a natural division defined by the $\omega_{2,0}$ quadrupole term from Eq.~(\ref{eq9}) will be used (the reason for this particular choice is explained in Sect.~\ref{results}). This quadrupole moment is proportional to $3\sin^2b-1$, and hence changes sign at $b_{\rm crit}=\arcsin\frac{1}{\sqrt{3}}\approx 35.2644^\circ$ and $-b_{\rm crit}\approx -35.2644^\circ$. This divides the sky into parts with $b\in(-b_{\rm crit},b_{\rm crit})$ (i.e., around the equator), and with $b<-b_{\rm crit}$ or $b>b_{\rm crit}$ (around the poles). To obtain the probability $p$ for the binomial test, one needs the area of a spherical cup with $b>b_{\rm crit}$. This is given by $S=2\pi(1-\sin b_{\rm crit})$, and $p=2S/(4\pi)\approx 0.4227$.

For benchmark testing, 10,000 MC realisations are generated according to Eq.~(\ref{eq2}) for short (278 points), long (1387 points), and a joined sample of 1669 points. For each realisation, the number of points with $b>b_{\rm crit}$ or $b<-b_{\rm crit}$ is counted and denoted $n_1$. Similarly, the number of points with $b\in(-b_{\rm crit},b_{\rm crit})$ is denoted $n_2$. The observed fraction is then $p_{\rm obs}=n_1/(n_1+n_2)=n_1/N$. Finally, the probability $P(N,n_1,p_{\rm obs})$ is computed according to Eq.~(\ref{eq12}), and compared with the significance level $\alpha=0.05$. If it exceeds $\alpha$, then that particular realisation passes the binomial test.

For all three sample sizes, more than 95\% of MC realisations passed the binomial test: 95.54\% for short GRBs, 95.45\% for long GRBs, and 95.01\% for a joined sample of all GRBs. This leads to a conclusion that a binomial test for the detection of anisotropy is trustworthy enough to use it for the observed GRB angular distribution.
\subsection{Two point angular correlation function (2pACF)}
\label{m9}
The 2pACF is a statistic used to characterize the clustering of objects on the sky, dependent on the angular scale $\theta$ and denoted $w(\theta)$. It is the projection of the spatial function on the sky and is defined by the joint probability $\delta P$ of finding two objects within the elements of solid angles $\delta\Omega_1$ and $\delta\Omega_2$, and separated by an angle $\theta$ with respect to that expected for an isotropic distribution \citep{peebles},
\begin{equation}
\delta P=m^2[1+w(\theta)]\delta\Omega_1\delta\Omega_2,
\label{eq13}
\end{equation}
where $m$ is the mean surface density of objects. If the distribution is isotropic, then $w(\theta)$ is equal to zero.

Among a number of estimators of the 2pACF, herein the \citet{landy} estimator is used, as it is one of the commonest nowadays. For its evaluation, one has to compute the distances between all $N$ points in the sample (data--data distances, $DD$; there are $N(N-1)/2$ such distances). Next, a so-called random catalog with $N_r\gg N$ is generated according to Eq.~(\ref{eq2}), and the distances between points in this catalog are computed (random--random, $RR$; there are $N_r(N_r-1)/2$ such distances), as well as the distances between the data and the random catalog (data--random, $DR$; there are $NN_r$ such distances). Herein, $N_r=10^5$ is used for all sample sizes. Finally, the distances are binned into bins with a given width ($1^\circ$ in this paper). The estimator has the form
\begin{equation}
w(\theta)=\frac{\braket{DD}}{\braket{RR}}-2\frac{\braket{DR}}{\braket{RR}}+1,
\label{eq14}
\end{equation}
where the triangular brackets denote the normalized mean. In full form, the 2pACF estimator is given by
\begin{equation}
w(\theta)=\frac{N_r(N_r-1)}{N(N-1)}\frac{DD}{RR}-\frac{N_r-1}{N}\frac{DR}{RR}+1,
\label{eq15}
\end{equation}
where $DD=DD(\theta)$, $DR=DR(\theta)$, and $RR=RR(\theta)$ are the numbers of pairs in the subsequent bins corresponding to the angular separations $\theta$, i.e., the counts in the range $(\theta,\theta+1^\circ)$. It is essential to note that the simulated sources in the random catalog have to be identical for $RR$ and $DR$.

Because it is extremely unlikely to achieve exactly $w(\theta)=0$ for any $\theta$, in order to estimate the error of the 2pACF for a given angle, a bootstrap method is applied \citep{efron1,efron2,efron3}. The procedure described above is repeated 100 times for different random catalogs. For each bootstrap realisation, the 2pACF is calculated, and the resulting errors are the standard deviations of the 100 values obtained for each angle $\theta$. The range of $\theta$ is from $0^\circ$ to $179^\circ$.

\subsection{2pACF benchmark testing}
\label{m10}
Because the 2pACF is computationally more expensive than the previous methods, and due to the fact that an MC method in the form of a bootstrap is embedded in the algorithm itself, for benchmark testing for each of the three sample sizes under consideration (278, 1387, and 1669 isotropic points generated according to Eq.~(\ref{eq2}), corresponding to short, long and all GRBs, respectively) one run of the algorithm is performed\footnote{Using a parallel {\sc C++} programme; available on request.}. The results in graphical form are displayed in Fig.~\ref{fig3}. The displayed values are the mean, and the errors are given by the standard deviation of the 100 bootstrap realisations. In Sect.~\ref{results}, the results of a 2pACF applied to the observed GRB samples will be compared with this figure.
\begin{figure}
\centering
\includegraphics[width=\columnwidth]{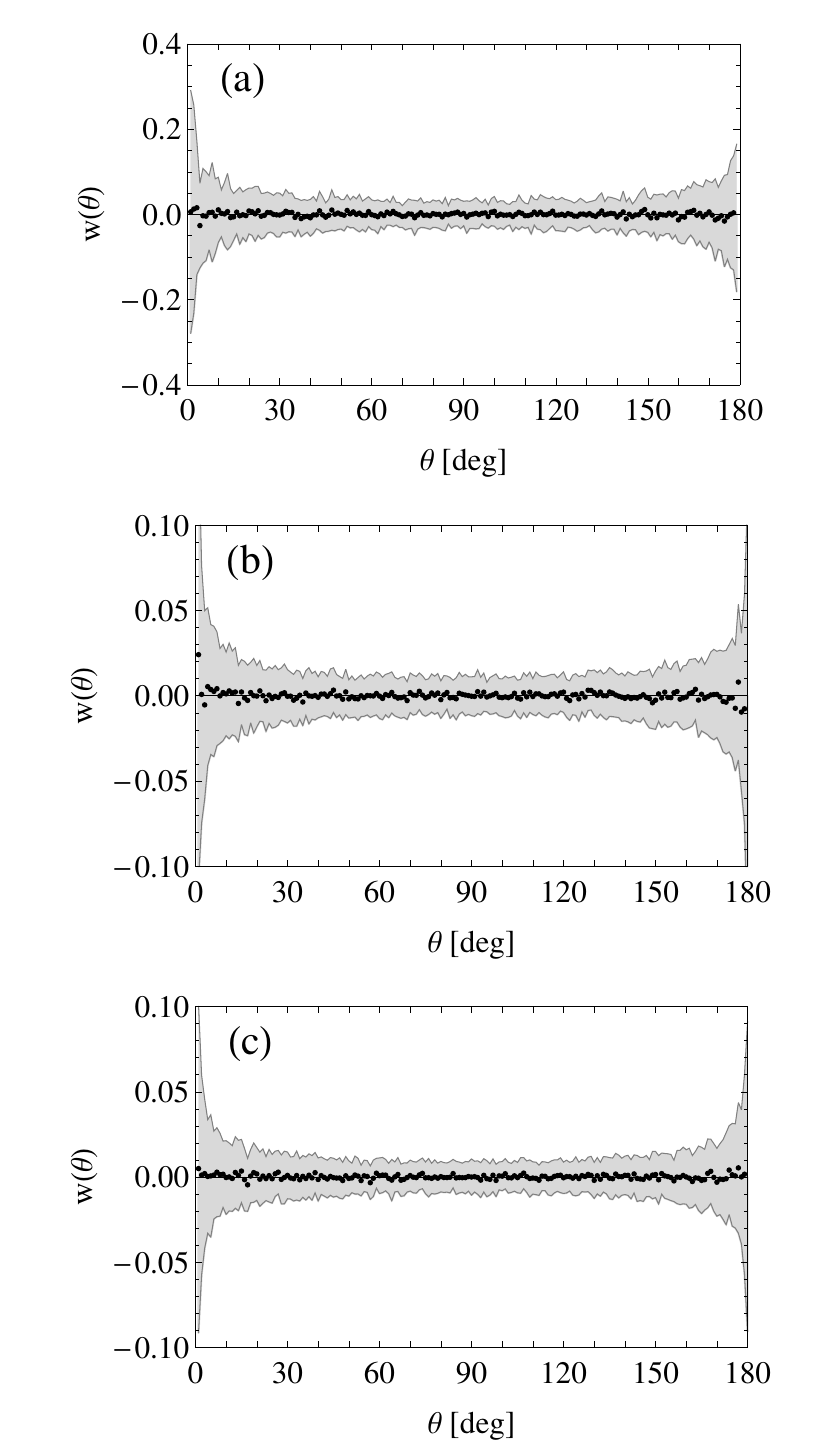}
\caption{The $w(\theta)$ function of a random realisation for (a) short, (b) long, and (c) all GRBs in the 2pACF benchmark testing. The shaded area marks the standard deviation of the values obtained. Its shape is similar in all three cases.}
\label{fig3}
\end{figure}
\section{Results}
\label{results}

\subsection{NNA}
\label{r1}
The NNA is performed on the samples of short, long and all GRBs as described in Sect.~\ref{m1} and \ref{m2}. First, the distributions of the first ($L=1$) and second ($L=2$) nearest neighbour distances are computed for each sample. For short GRBs, the mode for $L=1$ is at $4^\circ-6^\circ$, and for $L=2$ it lies between $6^\circ-8^\circ$. For long GRBs, these modes are located at $2^\circ-3^\circ$ and $4^\circ-5^\circ$ for the first and second nearest neighbour, respectively. Next, based on these angular distance distributions, the empirical CDFs are computed and compared with the theoretical CDF given by Eq.~(\ref{eq5}) and (\ref{eq6}). Both the empirical and theoretical CDFs are shown in Fig.~\ref{fig4},
\begin{figure*}
\centering
\includegraphics[width=\textwidth]{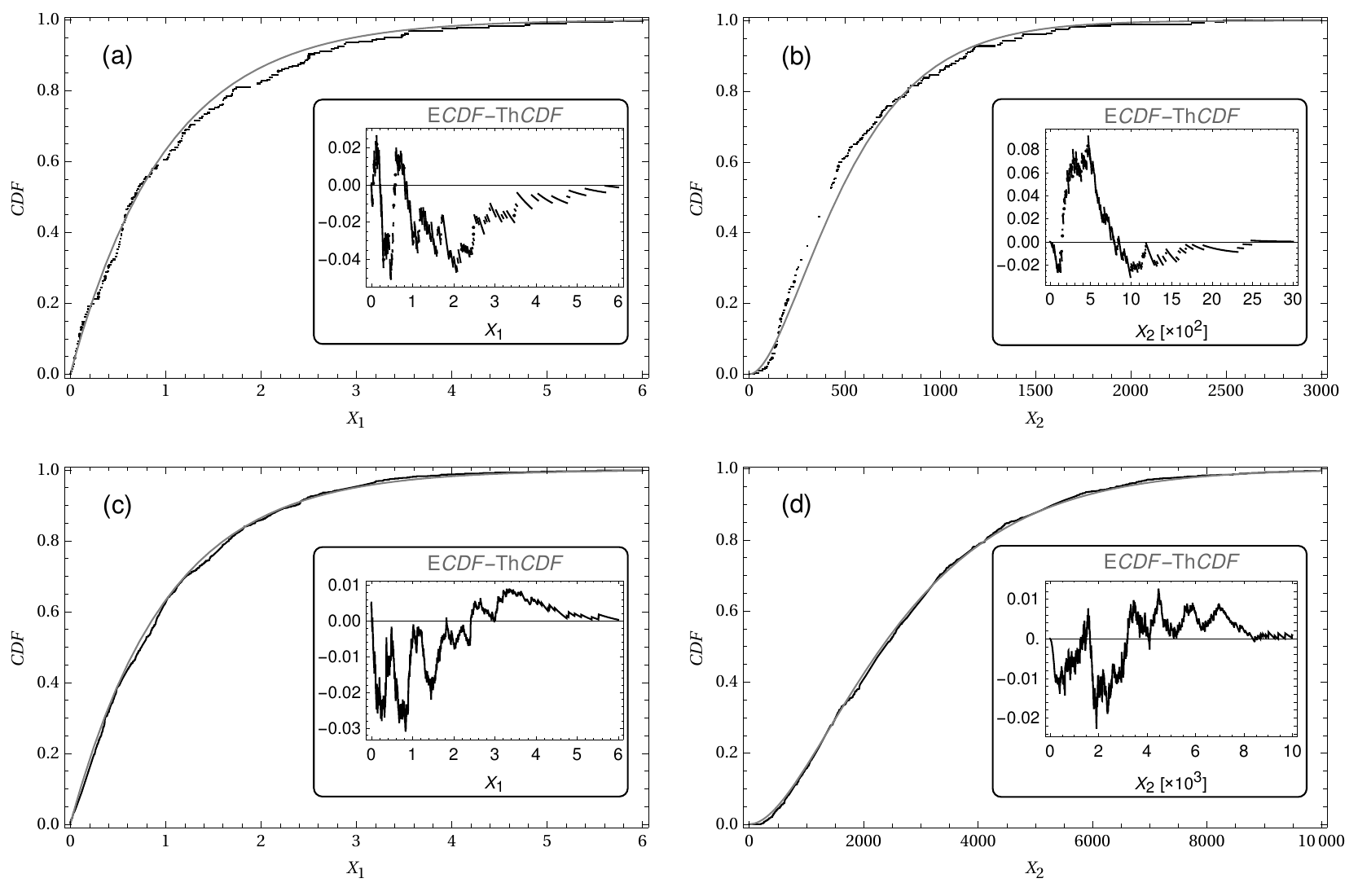}
\caption{The Kolmogorov-Smirnov test for the (a) first and (b) second nearest neighbour for short GRBs, and the (c) first and (d) second nearest neighbour for long GRBs. The black curve (the ragged one) is the empirical CDF, and the grey (the smooth) one is the theoretical CDF. The insets show the difference between the empirical and theoretical CDFs.}
\label{fig4}
\end{figure*}
where the insets show the absolute difference between the CDFs for each case. The plots for a joint sample of all GRBs are not very different from the ones for long GRBs, therefore are not displayed. Finally, the Kolmogorov-Smirnov test is performed, and its results are gathered in Table~\ref{tbl4}. The null hypothesis that the GRBs are distributed isotropically is rejected in the case of short GRBs based on the test applied to the second nearest neighbour ($L=2$) angular distance distribution. The biggest difference between the empirical and theoretical CDFs is at $X_2\approx 500$, which corresponds to an angular scale of $\approx 9^\circ$. The long and all GRBs are distributed isotropically according to the NNA.
\begin{table}
\begin{center}
\caption{The results of the Kolmogorov-Smirnov test for short, long and all GRBs in the NNA.}
\label{tbl4}
\begin{tabular}{cccccc}
\hline
Sample & $L$ & $N$ & $D_{\rm crit}$ & $D$ & $p$-value \\
\hline
\multirow{2}{*}{short} & 1 & \multirow{2}{*}{278}  & \multirow{2}{*}{0.0815} & 0.0520 & 0.4258 \\
					   & 2 & 					   & 					     & 0.0915 & 0.0178 \\
\multirow{2}{*}{long}  & 1 & \multirow{2}{*}{1387} & \multirow{2}{*}{0.0365} & 0.0311 & 0.1337 \\
					   & 2 & 					   & 					     & 0.0231 & 0.4417 \\
\multirow{2}{*}{all}   & 1 & \multirow{2}{*}{1669} & \multirow{2}{*}{0.0332} & 0.0243 & 0.2722 \\
					   & 2 & 					   & 					     & 0.0173 & 0.6962 \\
\hline
\end{tabular}
\end{center}
\end{table}

\subsection{Fractal dimension}
\label{r2}
The fractal dimension of short, long and all GRBs was computed as described in Sect.~\ref{m3} and \ref{m4}. The resulting $d_C$ values are gathered in Table~\ref{tbl5}. The correlation dimension for long and all GRBs is close to the expected value of $d_C=2$, but for short GRBs it departs rather strongly, achieving a value as small as $d_C=1.9051$. A one-sided and two-sided $p$-values are computed as defined in Sect.~\ref{m4}, and are the probabilities of finding, among a number of MC realisations of isotropic datasets, a dimension deviating from $d_C=2$ at least as much as the observed correlation dimension. For short and long GRBs, the one- and two-sided $p$-values are equal to each other (this means that there were no $d_C>2$ deviating from the expected value as much as the observed dimension), and for the sample of all GRBs they take different values. For short GRBs the $p$-value is small enough to doubt whether their celestial distribution is isotropic (it is equal to the significance level $\alpha$); most likely it is anisotropic. For long and all GRBs, the $p$-values are much greater than $\alpha$, hence with high confidence they are distributed isotropically on the sky.
\begin{table}
\begin{center}
\caption{Correlation dimensions of the angular distributions of short, long and all GRBs.}
\label{tbl5}
\begin{tabular}{ccccc}
\hline
Sample & $d_C$ & \begin{tabular}[c]{@{}c@{}}One-sided\\$p$-value\end{tabular} & \begin{tabular}[c]{@{}c@{}}Two-sided\\$p$-value\end{tabular} \\
\hline
short & 1.9051 & 0.05 & 0.05 \\
long  & 1.9941 & 0.58 & 0.58 \\
all   & 2.0108 & 0.50 & 0.61 \\
\hline
\end{tabular}
\end{center}
\end{table}

It is essential to note that long GRBs are about five times as numerous as short ones, hence the {\it contamination} of the dataset of all GRBs by short ones is negligible, and the isotropic distribution of long GRBs is dominating. This interpretation is consistent with the one-sided $p$-values---for long GRBs they are higher than for all GRBs, which means that the inclusion of short GRBs diminishes slightly the level of isotropy.
\subsection{Dipole-quadrupole test}
\label{r3}
The test is performed as described in Sect.~\ref{m5} and \ref{m6} for short, long and all GRBs. The coefficients $\omega_{i,j}$ were calculated according to Eq.~(\ref{eq10}), and a Student $t$ test was applied to verify whether they are equal to zero. For long and all GRBs, the null hypothesis that the coefficients are equal to zero cannot be rejected on the significance level $\alpha=0.05$, hence their angular distribution is isotropic. For a sample of short GRBs, only the $\omega_{2,0}$ coefficient did not pass the $t$ test, hence their sky distribution shows evidence for anisotropy, what is consistent with the results of the previous tests. However, due to a high percentage of false detections among the set of MC generated isotropic samples (compare with Sect.~\ref{m6}), the dipole-quadrupole test itself is not enough to claim anisotropy. Nevertheless, it is in agreement with the results of the previous tests.

\subsection{Binomial test}
\label{r4}
Based on the results of the dipole-quadrupole test, where only the $\omega_{2,0}$ coefficient is statistically different from zero in the case of short GRBs, the term corresponding to $\omega_{2,0}$ in Eq.~(\ref{eq9}), being proportional to $3\sin^2 b-1$, is used to divide the sky into two disjoint regions. As described in Sect.~\ref{m7} and \ref{m8}, the sky is divided into two parts: one around the equator, with $b\in(-b_{\rm crit},b_{\rm crit})$, and the second region is characterized by $b<-b_{\rm crit}$ or $b>b_{\rm crit}$, where $b_{\rm crit}\approx 35.2644^\circ$ (with the number of GRBs around both poles denoted with $n_1$).

For short, long and all GRBs, the number of objects in each of the region is counted straightforwardly, and the probability for the binomial test is $p_{\rm obs}=n_1/N$. Finally, the probability $P(N,n_1,p_{\rm obs})$ is calculated according to Eq.~(\ref{eq12}). The results are gathered in Table~\ref{tbl6}. The probability is very low for short GRBs, indicating anisotropy, and well above the significance level $\alpha=0.05$ for long and all GRBs. Again, the isotropy is weakened when the sample of long GRBs is contaminated with short ones, which add a level of anisotropy to the data. However, due to the fact that short GRBs are five times less numerous than long ones, the joint sample of all GRBs still passes the binomial test for isotropy. The results of the binomial test are again in agreement with the previous tests.
\begin{table}
\begin{center}
\caption{Results of the binomial test for short, long and all GRBs. The expected value of $n_1$ is equal to $Np$, with $p=0.4227$.}
\label{tbl6}
\begin{tabular}{ccccccc}
\hline
Sample & $N$ & $Np$ & $n_1$ & $p_{\rm obs}$ & $P(N,n_1,p_{\rm obs})$ \\
\hline
short & 278  & 118 & 137 & 0.4928 & 0.0150 \\
long  & 1387 & 586 & 589 & 0.4247 & 0.8705 \\
all   & 1669 & 705 & 729 & 0.4368 & 0.2343 \\
\hline
\end{tabular}
\end{center}
\end{table}
\subsection{2pACF}
\label{r5}
The 2pACF is applied to the observed celestial distribution of GRBs as described in Sect.~\ref{m9} and \ref{m10}, in the range $\theta\in[0^\circ,179^\circ]$, with an increament of $1^\circ$. The errors of $w(\theta)$ are estimated using the bootstrap method. The results, displayed in graphical form in Fig.~\ref{fig5}, are a somewhat noisy estimate.
\begin{figure}
\centering
\includegraphics[width=\columnwidth]{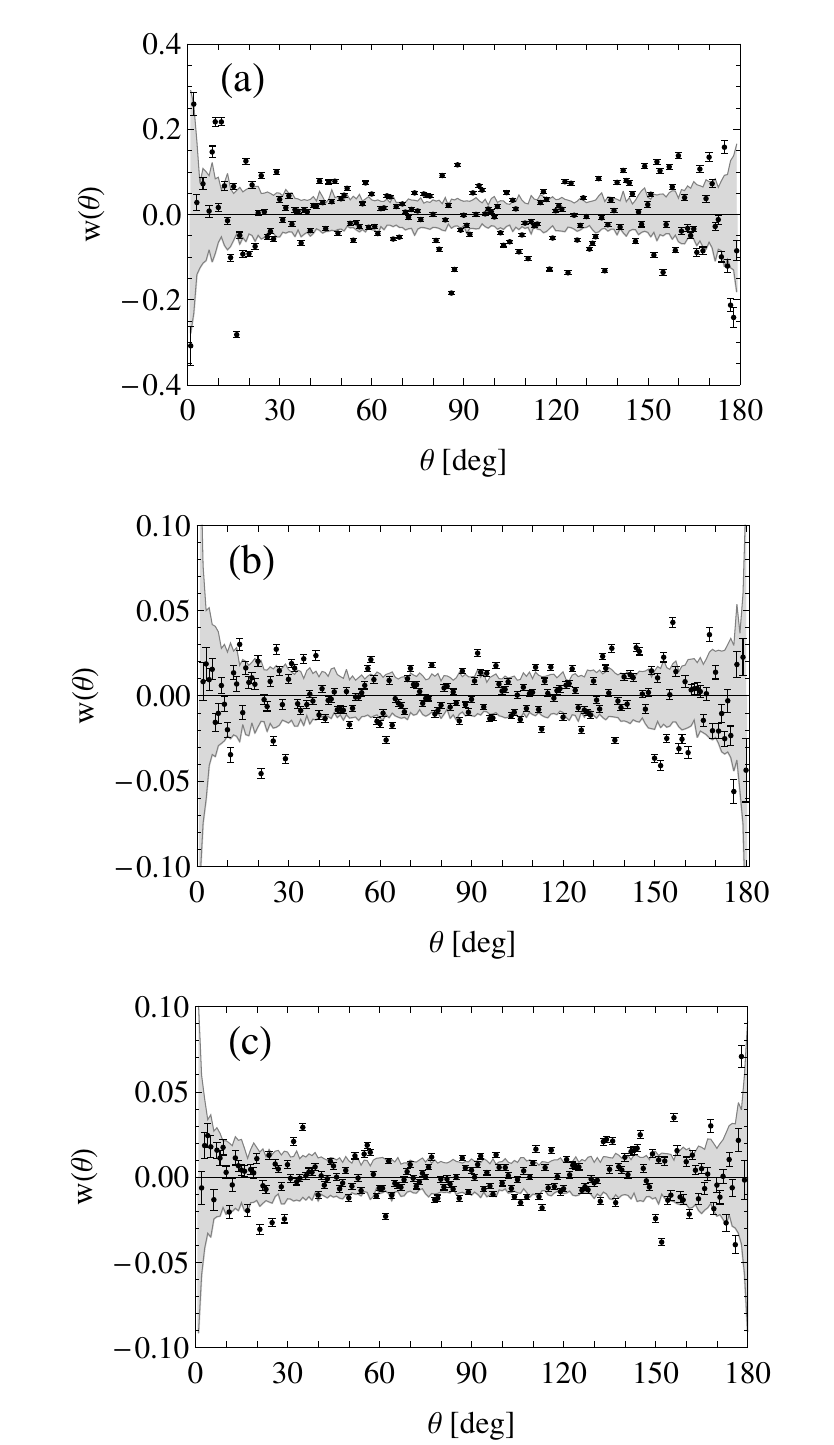}
\caption{The 2pACF for the observed (a) short, (b) long, and (c) all GRBs. The shaded area is the same standard deviation as in Fig.~\ref{fig3}, and it denotes the region in which a computed $w(\theta)$ value, if it falls in, may be still regarded as being statistically equal to zero.}
\label{fig5}
\end{figure}
The uncertainties of the 2pACF, marked with the error bars, do not allow to consider most of the $w(\theta)$ values to be equal to zero. Moreover, a significant number of the values obtained do not fall into the shaded area, which marks the region where the $w(\theta)$ might still be considered as a statistical zero, based on the benchmark testing from Sect.~\ref{m10}. What can be inferred from the figure, is that for short GRBs there are more $w(\theta)$ values that are outside the shaded area than there are in case of long and all GRBs. Therefore, one may say that the level of anisotropy is higher for short GRBs than for long ones.

While this finding is consistent with all the previous tests, as a stand-alone method it would not give a constructive conclusion. This outcome might be driven by the smallness of the examined sample of GRBs. Nevertheless, a qualitative statement that short GRBs are distributed anisotropically, is corroborated by the 2pACF method. This method, however, allows to put forward only a weaker, compared to claiming isotropy, statement that long GRBs are not as anisotropic as short ones. On the other hand, taking into account also the results of former tests, it can be said with a high level of confidence that the long GRBs are distributed isotropically (or at least almost isotropically) on the sky.

\section{Discussion}
\label{disc}
The significance level in the Kolmogorov-Smirnov test (Sect.~\ref{r1}) increases with $D\sqrt{N}$ \citep{meszaros03}. This means that it would not be surprising to find anisotropy among long GRBs and isotropy among short ones, as the former is about five times as numerous as the latter. In reality, the situation is exactly the opposite. This suggests that the isotropy found in the long GRBs, and the anisotropy among the short ones, is a reliable detection that does not arise simply due to the size $N$ of the samples. The discrepancy between the theoretical isotropic distribution and the actual distribution of short GRBs is greatest at angular scales $\theta\sim 9^\circ$, where the empirical CDF exceeds the theoretical one, indicating clustering at this scale. On the other hand, based on the benchmark testing (Sect.~\ref{m2}) the probability of a false anisotropy as a result of the NNA method is as high as 0.283, which is certainly not a level that allows to claim detection. Nevertheless, the results obtained hint toward rejection of isotropy in case of short GRBs and suggests that long ones are distributed isotropically on the sky. As a standalone test, however, this does not lead to reliable conclusions, hence other test are performed as well.

The correlation dimension obtained for short GRBs, compared with the results of the benchmark testing for this method (Sect.~\ref{m4} and \ref{r2}), indicates that anisotropy is indeed a reliable detection. The smallness of the sample may be seen as an obstacle in reliably estimating the correlation dimension. However, the probability of finding, among 1000 random MC simulated datasets with only $N=278$, a fractal dimension deviating from the expected $d_C=2$ at least as much as the observed value is only 5\%, comparable to the sigificance level $\alpha$. Hence, it is rather unlikely that the obtained $d_C$ is a chance occurence, and so the celestial distribution of short GRBs is anisotropic with a probability of 95\% based on the fractal dimension analysis.

The results of the dipole-quadrupole test (Sect.~\ref{r3}) are not trivial to interpret. All $\omega_{i,j}$ coefficients for the long and all GRBs were found to be statistically equal to zero, but the $\omega_{2,0}$ coefficient appeared to be the only non-zero one in case of short GRBs. This coefficient, being proportional to $3\sin^2b-1$, naturally divides the sky into a region around the equator, i.e. $b\in(-b_{\rm crit},b_{\rm crit})$, and around the poles, i.e. $b<-b_{\rm crit}$ or $b>b_{\rm crit}$, where $b_{\rm crit}\approx 35.2644^\circ$. The region around the equator might be ascribed to our Galaxy. However, even if one argues that this means that short GRBs might be of Galactic origin, the known redshifts strongly contradict this statement: while indeed short GRBs are characterised by redshifts systematically smaller than long GRBs, they tend to gather at a median redshift $\tilde{z}=0.78$, corresponding to a comoving distance of $2.8\,{\rm Gpc}$, which is a distance far away from the Galaxy (the median redshift for long GRBs is $\tilde{z}\approx 2$; see \citealt{zitouni,tarnopolski16b,tarnopolski16c}). Moreover, the existence of a non-zero $\omega_{2,0}$ quadrupole term with all the dipole terms being equal to zero is a strange behaviour, that does not allow to ascribe the source of GRBs to the Galaxy \citep{balazs}. Another way of explaining this phenomenon is to note that the Galaxy may play a role in the detection of GRBs by means of the {\it Fermi}/GBM sky-exposure function. However, this should not be the case, as i) the orbital period of {\it Fermi} is about 96 minutes, and the examined data sample covers almost 7 years, and ii) even if the sky exposure was not uniform, the anisotropy, as an instrumental effect, should be also visible in the distribution of long GRBs. Potentially it could be even stronger, as there are five times more long GRBs than short ones in the dataset, but the dipole-quadrupole test (as well as the other tests applied herein) yielded isotropy as a result. One last plea may be that the probability of obtaining a false-anisotropy is high, about 33\%, and that this might be the case here. Nevertheless, as was argued in Sect.~\ref{m6}, this test is a part of a compund of five tests, all of which are consistent with each other.

The argumentation for the binomial test (Sect.~\ref{r4}) is similar to the one for the dipole-quadrupole test. Despite an excessive number of short GRBs observed around the poles ($n_1=137$ versus the expected number of $Np=118$), the observed number of long GRBs $n_1=589$ is almost the same as the expected $Np=586$. Combining this with the non-zero quadrupole term corresponding to $\omega_{2,0}$ would suggest again an explanation that the short GRBs are of Galactic origin, which is extremely unlikey due to the known distribution of redshifts. Judging from the dipole-quadrupole and binomial tests alone, the Galaxy may still be the cause of the anisotropy, although after analysing the outcomes of other tests it seems unlikely, especially in light of isotropy among long GRBs.

The 2pACF (Sect.~\ref{r5}) is sensitive to bigger angular scales than NNA, but should detect anisotropy for the same instances (however, the opposite does not have to be true). The general picture, as displayed in Fig.~\ref{fig5}, is a bit puzzling due to a large scatter of the points in the $w(\theta)$ plot. However, qualitatively speaking, it is clear that the scatter is greater for short GRBs than for long ones, indicating a higher level of anisotropy. Unfortunately, in case of long GRBs a significant number of points lies outside the shaded area marking the region where the 2pACF corresponding to an isotropic random MC simulation would be expected to be. This allows at most to claim that the angular distribution of long GRBs is not as anisotropic as that of short GRBs. To quantify this claim, the following procedure is employed.

First, as seen in Fig.~\ref{fig5}, the absolute difference between the $w(\theta)$ and the boundary of the shaded area (the standard deviation) depends on $N$, hence the 2pACFs of long and all GRBs are normalized to the one for short GRBs. If one denotes by $s_{\rm sample}(\theta)$ the values of the standard deviation corresponding to an angle $\theta$, then the following function, being the sum of squared distances: $f(\sigma_{\rm sample})=\sum_{\theta=0}^{179}\left[ s_{\rm short}(\theta)-\sigma_{\rm sample} s_{\rm sample}(\theta) \right]^2$, is minimised with $\sigma_{\rm long}=2.5855$ and $\sigma_{\rm all}=3.2938$. Next, only $w(\theta)$ values that are outside the normalised shaded area are taken into account. The sum of distances $\sum d_i$, the sum of squared distances $\sum d_i^2$, and a straightforward square root of the latter of those points from the boundary of that area are calculated. The results are gathered in Table~\ref{tbl7}.
\begin{table}
\begin{center}
\caption{Establishing the relative strength of the isotropy detected with the 2pACF. See text for details.}
\label{tbl7}
\begin{tabular}{cccc}
\hline
 & short & long & all \\
\hline
$\sum d_i$ 						& 4.77 & 1.84 & 1.28 \\
$\sum d_i^2$				    & 0.69 & 0.30 & 0.07 \\
$\left(\sum d_i^2\right)^{1/2}$ & 0.83 & 0.55 & 0.26 \\
\hline
\end{tabular}
\end{center}
\end{table}

While these numbers are of unknown precise significance, they support the claim that the long GRBs are distributed more isotropically than short ones. In this light, the results of the 2pACF are consistent with the results of the previous tests. It is a bit surprising that the level of isotropy is smallest for the sample of all GRBs (all three deviation measures are smallest for this sample). One possible explanation might be that, as pointed above, the 2pACF is sensitive to bigger angular scales, where anisotropy is unlikely to be present for the sample of all GRBs. In combination with the fact that this is the biggest sample among the examined ones, it is possible that the influence of short GRBs on the overall anisotropy is smaller than statistical effects related to the sample size.

The fractal dimension, dipole-quadrupole, and binomial tests give an answer to the question whether GRBs are distributed isotropically, and qualitatively establish the strength of the anisotropy detected among short GRBs. The NNA, being sensitive to small angular scales (a few degrees), allows to expect an anisotropy at scales $\theta\approx 9^\circ$. This is in agreement with predictions from cosmology: the large-scale structure of the Universe is anisotropic at $z\lesssim 0.1$, or on scales $\lesssim 400\,{\rm Mpc}$.\footnote{The Sloan Great Wall \citep{gott} is an example of a huge structure that extends on that scale.} As most short GRBs with known $z$ have their redshifts below $z\sim 1$, this sets the scale above which the Universe is expected to be isotropic, i.e. the scale of averaging \citep{meszaros5}, to $\sim 3.4\,{\rm Gpc}$. As the GRBs are located in galaxies, they are expected to trace the large-scale distribution of matter in the Universe, and hence clustering of short GRBs is nothing unexpected. Nevertheless, due to an objective smallness of the GRB sample\footnote{Compare the number of known GRBs, aproximately 3000, to the number of galaxies in the SDSS, which exceeds one million \citep{sdss}.}, and the fact that they occur in galaxies very distant to each other\footnote{A GRB is expected to occur in a bright galaxy once per several million years \citep{loeb}, hence it would be uncommon to observe nearby events during the two decades of GRB observations.}, the relation between the lack of correlations between long GRBs (i.e., their isotropic distribution) and large-scale structures that might potentially contradict the cosmological principle is doubtful. On the other hand, the $z$ distribution of long BATSE GRBs was found to be likely proportional to the star formation rate, and while still plausible in case of short GRBs, the proportionality is less prominent \citep{meszaros2006}.

It is essential to emphasize that giant clusters consisting of quasars \citep[large quasar groups, LQGs;][]{clowes1,clowes2} or GRBs \citep{horvath14,balazs15,horvath15} have been discovered, with associated sizes as big as 1240 Mpc in case of the Huge-LQG \citep[HLQG;][]{clowes2}, or even 1720 Mpc in case of the giant GRB ring \citep{balazs15}. However, it was argued \citep{nadathur} that a single large structure does not violate the large-scale homogeneity. \citet{park} found that richness distribution of the richest 100, or the size distribution of the largest 100 quasar groups in SDSS DR7, are staistically equivalent with those of randomly distributed points, i.e. the large-scale structures like the HLQG can be found with high probability even if there is no physical clustering, and they do not challenge the homogeneous cosmological models. Moreover, it was shown \citep{li}, using the spatial two point correlation function (2pCF) on a sample of GRBs with known redshift, that the homogeneity scale is $\sim 11.7\,{\rm Gpc}$. Specifically, as mentioned in Sect.~\ref{intro}, the anisotropy detected among GRBs at $1.6<z<2.1$ \citep[31 GRBs;][]{horvath14} was not confirmed in a slightly larger sample in the same redshift range \citep[34 GRBs;][]{ukwatta}. The reasoning of \citet{nadathur}, initially applied to the HLQG \citep{clowes2}, and that of \citet{park}, originally applied to rich/large quasar groups, are also valid for other giant structures \citep[e.g.,][]{clowes1,balazs15} that were detected only by statistical inference, but their gravitational connection has not been verified. Hence, the existence of structures extending on cosmological distances does not need to contradict anisotropy, and vice versa: the anisotropy of short GRBs does not imply they form gravitationaly bounded groups. The difference between the above mentioned structures \citep{clowes1,clowes2,horvath14,balazs15,horvath15} and the analysis undertaken herein is that the angular GRB distribution examined in this paper is an all-sky one, with no redshift distribution taken explicitly into account. The sample of long GRBs, located in galaxies very distant from each other and with a significant range of redshifts, is not numerous enough to trace the population of their host galaxies, dark matter haloes, or other density fields. On the other hand, the detected anisotropy of short GRBs is a statistical feature of the whole available sample, and the existence of any large-scale structure is not claimed. Finally, it is worth to recall that the initials of Stephen Hawking, ``SH'', are clearly visible in the sky map of CMB from WMAP (\citealt{bennett}, Fig.~17 therein), which has an infinitesimal probability of occuring. Citing \citet{bennett}: ``It is clear that the combined selection of looking for initials, these particular initials, and their alignment and location are all a posteriori choices. (...) Low probability events are guaranteed to occur. The a posteriori assignment of a likelihood for a particular event detected, especially when the detection of that event is `optimized' for maximum effect by analysis choices, does not result in a fair unbiased assessment.''

The benchmark testing allows to establish the reliability of the anisotropy detection among the short GRBs examined herein. According to NNA (Sect.~\ref{m2}), the probability of falsely detecting an anisotropy among short GRBs is 0.283. In Sect.~\ref{r2}, the probability of a false anisotropy was estimated to be 0.05 when based on the correlation dimension. The dipole-quadrupole test in Sect.~\ref{m6} gave a probability of 0.337, and the binomial test in Sect.~\ref{m8} yielded a probability equal to 0.0446 of mistakenly classifying an isotropic dataset as an anisotropic one. Multiplying these probabilities, one gets 0.0002, or 0.02\%, of chances that the detection of anisotropy among short GRBs is a chance occurence and that the underlying population is intrinsically isotropic. This means that the significance of the anisotropy detected is 99.98\%. Taking into account the qualitative results of the 2pACF (Sect.~\ref{r5}), this probability is even higher. A crude estimation of the relative probability of anisotropy among short GRBs compared to long ones is done by taking ratios of the measure distances listed in Table~\ref{tbl7}. These ratios are comparable and equal to 0.386, 0.435, 0.663. Their average value, equal to 0.485, is taken to be the probability that the short GRBs are distributed anisotropically if the long ones have an isotropic distribution. The results of the other tests support the claim that long GRBs are isotropic, hence complementing the law of multiplication by the last value, the probability that short GRBs have an isotropic angular distribution, based on all five tests performed, is equal to 0.01\%, and the anisotropy is detected with a significance of 99.99\%. For the probability that the distribution of long GRBs is indeed isotropic, similar calculations give 30.68\%, without taking into account the 2pACF, which also favours the null hypothesis.

To conclude, it is important to note that the tests were performed for the sample of all GRBs only in order to further constrain the validity of each test, i.e., as it may be viewed as a set of long GRBs contaminated with a five times smaller number of short GRBs, one would expect the tests to give isotropy as a result, but with a smaller significance than for long GRBs alone. This was indeed the case, which confirms the validity and reliability of the tests applied. The sample of all GRBs, as being a composition of objects coming from different populations (mergers and non-mergers, i.e. short and long GRBs, respectively), does not have a real physical meaning, though.

\section{Conclusions}
\label{conc}
The angular distribution of 1669 GRBs observed by {\it Fermi}/GBM was examined. Various statistical tests (NNA, fractal dimension, dipole-quadrupole test, binomial test, and 2pACF) were conducted on a sample of 278 short GRBs, 1387 long ones, and a joined sample of all 1669 events (including four events with durations $T_{90}$ not established). The outcomes of all the tests consistently indicate that the population of short GRBs is distributed anisotropically on the sky, and long GRBs have an isotropic distribution. The reliability of each test was established based on a number of isotropically distributed on the sky sets of Monte Carlo generated points. The lowest ratio for sample size $N=278$ (corresponding to short GRBs), equal to 0.0446, of false-anisotropic detections was obtained for the binomial test. Moreover, the probability of achieving a fractal dimension as small as the observed one, is only 0.05. The ratios of incorrect classifications of an isotropic set by NNA and the dipole-quadrupole test are equal to 0.283 and 0.337, respectively. The multiplication rule of probability gives a net probability of 0.02\% that the four tests would all detect anisotropy in an isotropic dataset. Complementing this by a crude approximation based on the comparison of the 2pACF results for the short and long GRB samples (a conditional probability of 0.485 that short GRBs are anisotropic if the long ones are isotropic), the significance of the detected anisotropy is 99.99\%. Similar calculations for long GRBs give the probability of 30.68\% that the detected isotropic distribution is a real occurence (based on the NNA, fractal dimension, dipole-quadrupole and binomial tests, without the crude 2pACF estimate). The anisotropy among short GRBs does not violate the cosmological principle, and sets a scale of averaging above which the Universe is expected to be isotropic.

\section*{Acknowledgments}
The author is grateful to Tobiasz G\'orecki for fruitful discussions and help regarding the 2pACF, as well as for an efficient {\sc C++} programme for its computation.

\bsp

\label{lastpage}


\begin{thebibliography}{}
\bibitem[\protect\citeauthoryear{Ackermann et al.}{2013}]{ackermann} Ackermann, M., et al. 2013, ApJS, 209, 11
\bibitem[\protect\citeauthoryear{Alam et al.}{2015}]{sdss} Alam, S. 2015, ApJS, 219, 12
\bibitem[\protect\citeauthoryear{Alligood et al.}{2000}]{alligood} Alligood, K. T., Sauer, T. D., \& Yorke, J. A. 2000, Chaos. An Introduction to Dynamical Systems. Springer, Berlin
\bibitem[\protect\citeauthoryear{Atwood et al.}{2013}]{atwood} Atwood, W. B., et al. 2013, ApJ, 774, 76
\bibitem[\protect\citeauthoryear{Bennett et al.}{2011}]{bennett} Bennett, C. L., Hill, R. S., Hinshaw, G., et al. 2011, ApJS, 192, 17
\bibitem[\protect\citeauthoryear{Bal\'azs et al.}{1998}]{balazs} Bal\'azs, L. G., M\'esz\'aros, A., \& Horv\'ath I. 1998, A\&A, 339, 1
\bibitem[\protect\citeauthoryear{Bal\'azs et al.}{2003}]{balazs03} Bal\'azs, L. G., Bagoly, Z. S., Horv\'ath, I., M\'esz\'aros, A., \& M\'esz\'aros, P. 2003, A\&A, 401, 129
\bibitem[\protect\citeauthoryear{Bal\'azs et al.}{2015}]{balazs15} Bal\'azs, L. G., Bagoly, Z., Hakkila, J. E., et al. 2015, MNRAS, 452, 2236
\bibitem[\protect\citeauthoryear{Basak \& Rao}{2013}]{basak} Basak, R., \& Rao, A. R. 2013, MNRAS, 436, 3082
\bibitem[\protect\citeauthoryear{Berger}{2014}]{berger} Berger, E. 2014, ARA\&A, 52, 43
\bibitem[\protect\citeauthoryear{Bernui et al.}{2008}]{bernui} Bernui, A., Ferreira, I. S., \& Wuensche, C. A. 2008, ApJ, 673, 968
\bibitem[\protect\citeauthoryear{Bromberg et al.}{2013}]{bromberg} Bromberg, O., Nakar, E., Piran, T., \& Sari, R. 2013, ApJ, 764, 179
\bibitem[\protect\citeauthoryear{Carroll et al.}{1992}]{carroll} Carroll, S. M., Press, W. H., \& Turner, E. L. 1992, ARA\&A, 30, 499
\bibitem[\protect\citeauthoryear{Clowes et al.}{2011}]{clowes1} Clowes, R. G., et al. 2011, MNRAS, 419, 556
\bibitem[\protect\citeauthoryear{Clowes et al.}{2013}]{clowes2} Clowes, R. G., et al. 2013, MNRAS, 429, 2910
\bibitem[\protect\citeauthoryear{Colgate}{1974}]{colgate} Colgate, S. A. 1974, ApJ, 187, 333
\bibitem[\protect\citeauthoryear{Cucchiara et al.}{2011}]{cucchiara} Cucchiara, A., Levan, A. J., Fox, D. B., et al. 2011, ApJ, 736, 7
\bibitem[\protect\citeauthoryear{Dainotti et al.}{2016}]{dainotti} Dainotti, M., Del Vecchio, R., \& Tarnopolski., M. 2016, AdAst, in press (\href{https://arxiv.org/abs/1612.00618}{arXiv:1612.00618})
\bibitem[\protect\citeauthoryear{Efron}{1979}]{efron1} Efron, B. 1979, Ann. Stat., 7, 1
\bibitem[\protect\citeauthoryear{Efron}{1981}]{efron2} Efron, B. 1981, Biometrika, 68, 589
\bibitem[\protect\citeauthoryear{Efron \& Tibshirani}{1981}]{efron3} Efron, B., \& Tibshirani, R. J. 1994, An Introduction to the Bootstrap, CRC Press
\bibitem[\protect\citeauthoryear{Fishman \& Meegan}{1995}]{fishman} Fishman, G. J., \& Meegan, C. A. 1995, ARAA, 33, 415
\bibitem[\protect\citeauthoryear{Gao et al.}{2010}]{gao} Gao, H., Lu, Y., \& Zhang, S. N. 2010, ApJ, 717, 268
\bibitem[\protect\citeauthoryear{Gehrels \& Razzaque}{2013}]{gehrels2} Gehrels, N., \& Razzaque, S. 2013, Front. Phys., 8, 661
\bibitem[\protect\citeauthoryear{Gott et al.}{2005}]{gott} Gott, III, J. R., Juri\'c, M., Schlegel, D., et al. 2005, ApJ, 624, 463
\bibitem[\protect\citeauthoryear{Grassberger \& Procaccia}{1983}]{grassberger} Grassberger, P., \& Procaccia, I. 1983, Physica D, 9, 189
\bibitem[\protect\citeauthoryear{Grassberger}{1986}]{grassberger2} Grassberger, P. 1986, Nature, 323, 609
\bibitem[\protect\citeauthoryear{Gruber}{2012}]{gruber2} Gruber, D. 2012, in Proceedings of the Gamma-Ray Bursts 2012 Conference (GRB 2012). Munich, Germany, 007
\bibitem[\protect\citeauthoryear{Gruber et al.}{2014}]{gruber} Gruber, D., et al. 2014, ApJS, 211, 12
\bibitem[\protect\citeauthoryear{Hamilton}{1993}]{hamilton} Hamilton, A. J. S. 1993, ApJ, 417, 19
\bibitem[\protect\citeauthoryear{Horv\'ath et al.}{1996}]{meszaros8} Horv\'ath, I., M\'esz\'aros, P., \& M\'esz\'aros, A. 1996, ApJ, 470, 56
\bibitem[\protect\citeauthoryear{Horv\'ath}{1998}]{horvath98} Horv\'ath, I. 1998, ApJ, 508, 757
\bibitem[\protect\citeauthoryear{Horv\'ath}{2002}]{horvath02} Horv\'ath, I. 2002, A\&A, 392, 791
\bibitem[\protect\citeauthoryear{Horv\'ath}{2009}]{horvath09} Horv\'ath, I. 2009, Ap\&SS, 323, 83
\bibitem[\protect\citeauthoryear{Horv\'ath et al.}{2008}]{horvath08} Horv\'ath, I., Bal\'azs, L. G., Bagoly, Z., et al. 2008, A\&A, 489, L1
\bibitem[\protect\citeauthoryear{Horv\'ath et al.}{2010}]{horvath10} Horv\'ath, I., Bagoly, Z., Bal\'azs, L. G., et al. 2010, ApJ, 713, 552
\bibitem[\protect\citeauthoryear{Horv\'ath et al.}{2014}]{horvath14} Horv\'ath, I., Hakkila, J., \& Bagoly, Z. 2014, A\&A, 561, L12
\bibitem[\protect\citeauthoryear{Horv\'ath et al.}{2015}]{horvath15} Horv\'ath, I., Bagoly, Z., Hakkila, J., et al. 2015, A\&A, 584, A48
\bibitem[\protect\citeauthoryear{Huja et al.}{2009}]{huja} Huja, D., M\'esz\'aros, A., \& \v R\'{\i}pa, J. 2009, A\&A, 504, 67
\bibitem[\protect\citeauthoryear{Janiuk et al.}{2006}]{janiuk} Janiuk, A., Czerny, B., Moderski, R., et al. 2006, MNRAS, 365, 874
\bibitem[\protect\citeauthoryear{Kann et al.}{2011}]{kann} Kann, D. A., Klose, S., Zhang, B., et al. 2011, ApJ, 734, 96
\bibitem[\protect\citeauthoryear{Kendall \& Stuart}{1973}]{kendall} Kendall, M., \& Stuart, A. 1973, The Advanced Theory of Statistics. Griffin, London
\bibitem[\protect\citeauthoryear{Klebesadel et al.}{1973}]{klebesadel} Klebesadel, R. W., Strong, I. B., \& Olson, R. A. 1973, ApJ, 182, L85
\bibitem[\protect\citeauthoryear{Kouveliotou et al.}{1993}]{kouve} Kouveliotou, C., Meegan, C. A., Fishman, G. J., et al. 1993, Apj, 413, L101
\bibitem[\protect\citeauthoryear{Landy \& Szalay}{1993}]{landy} Landy, S. D., \& Szalay, A. S. 1993, ApJ, 412, 64
\bibitem[\protect\citeauthoryear{Li \& Lin}{2015}]{li} Li, M.-H., \& Lin, H.-N. 2015, A\&A, 582, A111
\bibitem[\protect\citeauthoryear{Li et al.}{2016}]{li2} Li, Y., Zhang, B., \& L\"u, H.-J. 2016, ApJS, 227, 7
\bibitem[\protect\citeauthoryear{Loeb}{1998}]{loeb} Loeb, A. 1998, ApJ, 503, L35
\bibitem[\protect\citeauthoryear{Magliocchetti et al.}{2003}]{maglio} Magliocchetti, M., Ghirlanda, G., \& Celotti, A. 2003, MNRAS, 343, 255
\bibitem[\protect\citeauthoryear{Mandelbrot}{1983}]{mandelbrot} Mandelbrot, B. 1983. The Fractal Geometry of Nature. W. H. Freeman and Company, New York
\bibitem[\protect\citeauthoryear{Massey}{1951}]{kolmogorov} Massey, F. J. Jr 1951, J. Am. Stat. Assoc., 46, 68
\bibitem[\protect\citeauthoryear{Mazets et al.}{1981}]{mazets} Mazets, E. P., et al. 1981, Ap\&SS, 80, 3
\bibitem[\protect\citeauthoryear{Meegan et al.}{1992}]{meegan92} Meegan, C. A., Fishman, G. J., Wilson, R. B., et al. 1992, Nature, 355, 143
\bibitem[\protect\citeauthoryear{Metzger et al.}{1997}]{metzger} Metgzer, M. R., Djorgovski, S. G., Kulkarni, S. R., Steidel, C. C., Adelberger, K. L., Frail, D. A., Costa, E., \& Frontera, F. 1997, Nature, 387, 878
\bibitem[\protect\citeauthoryear{M\'esz\'aros \& M\'esz\'aros}{1996}]{meszaros7} M\'esz\'aros, A., \& M\'esz\'aros, P. 1996, Apj, 466, 29
\bibitem[\protect\citeauthoryear{M\'esz\'aros}{1997}]{meszaros97} M\'esz\'aros, A. 1997, A\&A, 328, 1
\bibitem[\protect\citeauthoryear{M\'esz\'aros \& \v Sto\v cek}{2003}]{meszaros03} M\'esz\'aros, A., \& \v Sto\v cek, J. 2003, A\&A, 403, 443
\bibitem[\protect\citeauthoryear{M\'esz\'aros et al.}{2000a}]{meszaros4} M\'esz\'aros, A., Bagoly, Z., Horv\'ath, I., et al. 2000a, ApJ, 539, 98
\bibitem[\protect\citeauthoryear{M\'esz\'aros et al.}{2000b}]{meszaros3} M\'esz\'aros, A., Bagoly, Z., \& Vavrek, R., 2000b, A\&A, 354, 1
\bibitem[\protect\citeauthoryear{M\'esz\'aros et al.}{2006}]{meszaros2006} M\'esz\'aros, A., Bal\'azs, L. G., Bagoly, Z., et al. 2006, A\&A, 455, 785
\bibitem[\protect\citeauthoryear{M\'esz\'aros et al.}{2009a}]{meszaros} M\'esz\'aros, A., Bal\'azs, L. G., Bagoly, Z., et al. 2009a, AIP Conf. Proc., 1133, 483
\bibitem[\protect\citeauthoryear{M\'esz\'aros et al.}{2009b}]{meszaros5} M\'esz\'aros, A., Bal\'azs, L. G., Bagoly, Z., et al. 2009b, Balt. Astron., 18, 293
\bibitem[\protect\citeauthoryear{M\'esz\'aros \& M\'esz\'aros}{1995}]{meszaros6} M\'esz\'aros, P., \& M\'esz\'aros, A. 1995, ApJ, 449, 9
\bibitem[\protect\citeauthoryear{M\'esz\'aros \& Rees}{2015}]{rees} M\'esz\'aros, P., \& Rees, M. J. 2015, in Ashtekar, A., Berger, B., Isenberg, J., \& MacCallum, M. A. H., eds, General Relativity and Gravitation: A Centennial Perspective. Cambridge Univ. Press, Cambridge
\bibitem[\protect\citeauthoryear{Mukherjee et al.}{1998}]{mukh} Mukherjee, S., Feigelson, E. D., Jogesh Babu, G., et al. 1998, ApJ, 508, 314
\bibitem[\protect\citeauthoryear{Muccino et al.}{2013}]{muccino} Muccino, M., Ruffini, R., Bianco, C. L., et al. 2013, ApJ, 763, 125
\bibitem[\protect\citeauthoryear{Nadathur}{2013}]{nadathur} Nadathur, S. 2013, MNRAS, 434, 398
\bibitem[\protect\citeauthoryear{Nakar}{2007}]{nakar} Nakar, E. 2007, Phys. Rep., 442, 166
\bibitem[\protect\citeauthoryear{Ott}{2002}]{ott} Ott, E. 2002. Chaos in Dynamical Systems. Cambridge Univ. Press, Cambridge
\bibitem[\protect\citeauthoryear{Paczy\'nski}{1986}]{paczynski86} Paczy\'nski, B. 1986, ApJ, 308, L43
\bibitem[\protect\citeauthoryear{Paczy\'nski}{1991}]{paczynski91} Paczy\'nski, B. 1991, Acta Astron., 41, 257
\bibitem[\protect\citeauthoryear{Park et al.}{2015}]{park} Park, C., et al. 2015, JKAS, 48, 75
\bibitem[\protect\citeauthoryear{Peebles}{1980}]{peebles} Peebles, P. 1980. The Large-Scale Structure of the Universe. Princeton Univ. Press, Princeton
\bibitem[\protect\citeauthoryear{Planck Collaboration XLVI}{2016}]{planck} Planck Collaboration XLVI. 2016, A\&A, 596, A107
\bibitem[\protect\citeauthoryear{Racusin et al.}{2011}]{racusin} Racusin, J. L., et al. 2011, ApJ, 738, 138
\bibitem[\protect\citeauthoryear{R\'{\i}pa et al.}{2009}]{ripa1} R\'{\i}pa, J., M\'esz\'aros, A., Wigger, C., et al. 2009, A\&A, 498, 399
\bibitem[\protect\citeauthoryear{R\'{\i}pa et al.}{2012}]{ripa2} R\'{\i}pa, J., M\'esz\'aros, A., Veres, P., et al. 2012, ApJ, 756, 44
\bibitem[\protect\citeauthoryear{R\'{\i}pa et al.}{2016}]{ripa3} R\'{\i}pa, J., \& M\'esz\'aros, A. 2016, Ap\&SS, 361:370
\bibitem[\protect\citeauthoryear{Sachs}{1997}]{sachs} Sachs, L. 1997, Angewandte Statistik, Springer
\bibitem[\protect\citeauthoryear{Scott \& Tout}{1989}]{scott} Scott, D., \& Tout, C. A. 1989, MNRAS, 241, 109
\bibitem[\protect\citeauthoryear{\v Slechta \& M\'esz\'aros}{1997}]{slechta} \v Slechta, M., \& M\'esz\'aros, A. 1997, Ap\&SS, 249, 1
\bibitem[\protect\citeauthoryear{Student}{1908}]{student} Student 1908, Biometrika, 6, 1
\bibitem[\protect\citeauthoryear{Tarnopolski}{2014}]{tarnopolski14} Tarnopolski, M. 2014, Rom. Rep. Phys., 66, 907
\bibitem[\protect\citeauthoryear{Tarnopolski}{2015a}]{tarnopolski15a} Tarnopolski, M. 2015a, Ap\&SS, 359:20
\bibitem[\protect\citeauthoryear{Tarnopolski}{2015b}]{tarnopolski15b} Tarnopolski, M. 2015b, A\&A, 581, A29
\bibitem[\protect\citeauthoryear{Tarnopolski}{2016a}]{tarnopolski16a} Tarnopolski, M. 2016a, MNRAS, 458, 2024
\bibitem[\protect\citeauthoryear{Tarnopolski}{2016b}]{tarnopolski16b} Tarnopolski, M. 2016b, New Astron., 46, 54
\bibitem[\protect\citeauthoryear{Tarnopolski}{2016c}]{tarnopolski16c} Tarnopolski, M. 2016c, Ap\&SS, 361:125
\bibitem[\protect\citeauthoryear{Tegmark et al.}{1996}]{tegmark} Tegmark, M., Hartmann, D. H., Briggs, M. S., et al. 1996, ApJ, 468, 214
\bibitem[\protect\citeauthoryear{Ukwatta \& Wo\'zniak}{2015}]{ukwatta} Ukwatta, T. N., \& Wo\'zniak, P. R. 2015, MNRAS, 455, 703
\bibitem[\protect\citeauthoryear{Usov \& Chibisov}{1975}]{usov} Usov, V. V., \& Chibisov, G. V. 1975, SvA, 19, 115
\bibitem[\protect\citeauthoryear{Woosley \& Bloom}{2006}]{woosley} Woosley, S. E., \& Bloom, J. S. 2006, ARA\&A, 44, 507
\bibitem[\protect\citeauthoryear{Vavrek et al.}{2008}]{vavrek} Vavrek, R., Bal\'azs, L. G., M\'esz\'aros, A., et al. 2008, MNRAS, 391, 1741
\bibitem[\protect\citeauthoryear{Veres et al.}{2010}]{veres} Veres, P., Bagoly, Z., Horv\'ath, I., et al. 2010, ApJ, 725, 1955
\bibitem[\protect\citeauthoryear{von Kienlin et al.}{2014}]{kienlin} von Kienlin, A., et al. 2014, ApJS, 211, 13
\bibitem[\protect\citeauthoryear{Yadav et al.}{2010}]{yadav} Yadav, J. K., Bagla, J. S., \& Khandai, N. 2010, MNRAS, 405, 2009
\bibitem[\protect\citeauthoryear{Zhang}{2011}]{zhang4} Zhang, B. 2011, C. R. Phys., 12, 206
\bibitem[\protect\citeauthoryear{Zhang \& Choi}{2008}]{zhang2} Zhang, Z.-B., \& Choi, C.-S. 2008, A\&A, 484, 293
\bibitem[\protect\citeauthoryear{Zhang et al.}{2009}]{zhang5} Zhang B., Zhang B.-B., Virgili F. J., et al. 2009, ApJ, 703, 1696
\bibitem[\protect\citeauthoryear{Zhang et al.}{2012}]{zhang3} Zhang, F.-W., Shao, L., Yan, J.-Z., et al. 2012, ApJ, 750, 88
\bibitem[\protect\citeauthoryear{Zitouni et al.}{2015}]{zitouni} Zitouni, H., Guessoum, N., Azzam, W. J., et al. 2015, Ap\&SS, 357:7

\end{thebibliography}
\end{document}